\documentclass{aa}  

\usepackage{natbib}
\usepackage{graphicx}
\usepackage{txfonts}
\usepackage[colorlinks=true, allcolors=blue]{hyperref}

\usepackage{amsmath}

\begin{document}

   \title{The Main Evolutionary Pathways of Massive Hierarchical Triple Stars}


   \author{F. Kummer
          \inst{1}
          \and
          S. Toonen\inst{1}
          \and A. de Koter\inst{1,2}
          }

   \institute{Anton Pannekoek Institute for Astronomy, University of Amsterdam, Science Park 904, 1098 XH Amsterdam, Netherlands
              \and
              Institute of Astronomy, KU Leuven, Celestijnenlaan 200D, 3001 Leuven, Belgium\\ 
              \email{f.a.kummer@uva.nl}
             }

   \date{Received ... ; accepted ...}
 
  \abstract
   {So far, stellar population studies have mainly focused on the evolution of single and binary stars. Recent observations show that triple and higher order multiple star systems are ubiquitous in the local population, especially among massive stars. Introducing three-body dynamical effects can influence the evolution of an individual stellar system and hence affect the predicted rates of astrophysical sources that are a product of stellar evolution. Therefore, predictions of triple star evolution are necessary for a more complete understanding of the evolutionary behavior of stellar populations and their end products.}
   {We aim to constrain the main evolutionary pathways of massive hierarchical triple star systems and to quantify the effect of the third star on the evolution of the system.}
   {We model the massive triple star population by performing simulations of  triple star evolution with the TRES code, which combines stellar evolution with secular evolution of triple systems, and explore how robust the predictions of these simulations are under variations of uncertain initial conditions. 
   We focus on coeval, hierarchical stellar triples in pre-mass-transfer phases.}
  {Interactions are common among massive triple stars. The majority of systems (65-77\%) experience a phase of mass transfer in the inner binary, often with an unevolved donor star. This differs significantly from isolated binary evolution, where mass transfer is less frequent (52.3\% instead of 67\% for our fiducial model) and the donors are typically post-main sequence stars. Initial constraints for dynamical stability as well as eccentricity oscillations driven by the third body facilitate the occurrence of interactions, for instance mass transfer. The requirement of dynamical stability at formation places quite stringent constraints on allowed orbital properties, reducing uncertainties in  triple evolution that resort from these initial conditions. Ignoring three-body dynamics during evolution of non-interacting triples leads to triple compact-object systems with stronger eccentricity oscillations and thereby likely over-predicts the merger rate of compact objects in such systems.}    {}


   \maketitle
%

\section{Introduction}

Massive stars ($M\gtrsim8\rm{M}_{\odot}$) play an important role in the dynamical and chemical evolution of galaxies. Stellar feedback in the form of stellar winds, radiation pressure, photo-ionization and supernova (SN) explosions control star formation \citep[e.g.][]{hayward_how_2017} and enrich the interstellar medium with metals \citep[e.g.][]{burbidge_synthesis_1957, larson_effects_1974, larson_gas_1975}. Most massive stars are observed to have at least one stellar companion \citep[e.g.][]{evans_vlt-flames_2006,sana_southern_2014,kobulnicky_toward_2014, moe_mind_2017} and are expected to interact at some stage of their evolution \citep{sana_binary_2012}. These interacting systems can be the progenitors for a great variety of observed astrophysical sources, such as gravitational wave (GW) mergers \citep[e.g.][]{abbott_observation_2016}, X-ray binaries \citep[e.g.][]{verbunt_origin_1993,langer_presupernova_2012}, massive equivalents of Algol systems \citep[e.g.][]{budding_resolving_1989} and blue stragglers \citep[e.g.][]{perets_triple_2009}, and ultra-stripped supernovae \citep[e.g.][]{tauris_ultra-stripped_2013}. Mass transfer, specifically, can significantly impact the subsequent evolution of both the donor and accretor stars \citep[e.g.][]{eldridge_effect_2008,langer_presupernova_2012,laplace_different_2021}.

In the past, the vast majority of theoretical work in stellar evolution has focused on single- and binary stellar models. However, more recent advances in observational astronomy reveal that higher order multiples are prevalent, especially among the massive star population. Roughly 10\% of the solar-type star systems are triples \citep{tokovinin_binaries_2014, moe_mind_2017}, while for early B-type and O-type stars this fraction increases to over 30\% \citep{evans_vlt-flames_2006,moe_mind_2017}. Unsurprisingly, stellar products have been observed whose formation poses a challenge from the perspective of single- and binary evolution, but can more easily be explained with the inclusion of a third companion, e.g., low mass X-ray binaries containing a black hole \citep{podsiadlowski_formation_2003, naoz_formation_2016}, barium stars \citep{gao_stellar_2023}, cataclysmic variables with high accretion rates \citep{knigge_triple_2022} and binary stars with one magnetic component \citep{alecian_binamics_2014, schneider_rejuvenation_2016, schneider_stellar_2019, shultz_mobster_2019}. Therefore, complimentary population synthesis studies of triples are essential toward creating a more comprehensive picture of stellar evolution and interaction. 

Introducing a third companion to a binary adds complexity to the system. On top of stellar evolution and interactions, three-body dynamics has to be taken into account to accurately describe the evolution of a triple star system over time. The lowest order manifestation of these three-body interactions in a stable triple system are the Von Zeipel-Lidov-Kozai (ZLK) oscillations \citep{von_zeipel_sur_1910,lidov_evolution_1962,kozai_secular_1962}, which are periodic variations of the eccentricity of the orbit of the inner two stars and the inclination of the third star with respect to the orbital plane of the inner orbit. For an elaborate review on the effects of three-body dynamics, see \citet{naoz_eccentric_2016}. Recent population synthesis studies of triple stars have shown that increased eccentricities induced by three-body dynamical effects such as ZLK oscillations can lead to an enhanced degree of stellar interaction between the components of the inner binary compared to such interactions in isolated binary stars \citep{antonini_binary_2017, hamers_double_2019, toonen_evolution_2020, stegmann_evolution_2022, stegmann_binary_2022, hamers_statistical_2022, hamers_return_2022}.

In this paper, we aim to provide an inventory of the most common evolutionary channels within a population of massive triple stars and show how these predictions differ from standard binary evolution. To explore this, we perform simulations of massive hierarchical triple stars with a triple population synthesis code. We stop the computation of an individual stellar and orbital evolution when the system becomes dynamically unstable, unbound or at the onset of mass transfer, i.e., for this subset of evolutionary channels we do not continue until the final remnant stages. The main reasons for this are that the physics of mass transfer in a triple configuration is still poorly understood and awaits more detailed studies, and that it allows us to focus on the effects of a third companion on the initial orbital conditions and early evolution. 

In Section 2, we give a description of the triple population synthesis code and motivate our choice of initial model conditions. In Section 3, we present the predicted incidence rates of each evolutionary channel and discuss a few channels in more detail. In Section 4, we compare our results with similar studies and discuss some caveats. Finally, we give a summary of our findings in Section 5.


\section{Method}
\label{sect:method}


To investigate the most common evolutionary channels in massive hierarchical triple stars we use the triple population synthesis code TRES\footnote{This code is publicly available on: \url{https://github.com/amusecode/TRES}} \citep{toonen_evolution_2016}. This is a code that couples a single and binary stellar evolution code, in this case SeBa \citep{portegies1996population,toonen_supernova_2012}, which models stellar evolution, and a code that solves the orbital motions of three-body systems using the secular approximation \citep{toonen_evolution_2016}. In the secular approximation, the system is sufficiently hierarchical such that the tertiary star effectively orbits the center of mass of the inner two stars, forming the outer orbit. In these configurations, the energy of the outer orbit and the orbit of the inner two stars is separately conserved \citep{heggie_binary_1975}.
To model stellar evolution, SeBa makes use of analytic fitting formulae from \citet{hurley_comprehensive_2000, hurley_evolution_2002}, based on the single stellar evolution models produced by \citet{pols_approximate_1995}. These fitting formulae allow for rapid simulation of stellar evolution, essential for population synthesis studies. 

With TRES, we simulate ten sets of 25000 initially dynamically stable, coeval, massive hierarchical triple stars that are initialized at the zero age main sequence (ZAMS). These systems are then evolved for approximately a Hubble time (13.5 Gyr), unless the systems experience a phase of mass transfer, become dynamically unstable or unbind, after which the simulation is terminated. In a very small number of simulations ($\lesssim$ 0.05\%) the timescale for dynamical interactions is extremely short compared to the nuclear timescale; we decided to discontinue these models for practical reasons. For a similar number of systems, the secular code could not find an analytical solution for the orbital elements. These too were discarded. 

We will first discuss the adopted treatment of mass loss in stellar winds and our choices for the initial stellar and orbital properties of the primordial triple star population. Subsequently, we review a few physical implications of multiplicity that can affect the evolution of a system.

\subsection{Updated wind prescription}
By default, SeBa follows the line-driven wind mass-loss model for massive stars of \citet{vink_new_2000, vink_mass-loss_2001} in the metallicity range $0.03\:Z_{\odot}<Z<3\:Z_{\odot}$ and \citet{nieuwenhuijzen1990parametrization} in the range where the Vink models are not applicable. We have modified these prescriptions by reducing the mass-loss rates by a factor 3, following and generalising the findings of \citet{bjorklund_new_2021} for the implemented mass range. Additionally, we have assumed luminous blue variable mass-loss rates following \citet{belczynski_maximum_2010}, introducing a constant enhanced mass loss of $1.5\times10^{-4}\:\rm{M}_{\odot}/\rm{yr}$ for stars exceeding a luminosity of $6\times10^5\:\rm{L}_{\odot}$.  

\subsection{Initial parameters}
\label{sect: init_params}
To simplify the description of a hierarchical triple star, the system is split up into two subsystems: (i) The inner binary, that comprises the primary and secondary star with masses $m_{1}$ and $m_{2}$. These stars orbit each other with an inner semi-major axis $a_{\rm{in}}$, eccentricity $e_{\rm{in}}$ and argument of pericenter $g_{\rm{in}}$; (ii) The outer binary, that comprises the centre of mass of the inner binary and the tertiary star, with mass $m_{3}$. The outer binary has an outer semi-major axis $a_{\rm{out}}$, eccentricity $e_{\rm{out}}$ and argument of pericenter $g_{\rm{out}}$. We define a relative inclination angle $i_{\rm{rel}}$ between the orbital planes of the inner and outer orbit. 

The incidence rates of transients depend on the ZAMS conditions of a stellar population \citep[e.g.][]{abate_modelling_2015,de_mink_merger_2015,stevenson_wide_2022}. Therefore, we would ideally like to possess a complete understanding of how the initial stellar and orbital parameters of massive triples are distributed. Unfortunately, the observed sample size of massive stars is limited compared to low-mass stars and our understanding is fragmented, as certain parts of the parameter space (e.g., low mass ratios and wide orbits) are difficult to probe. Moreover, observations suggest there is a degeneracy in constraining certain parameter combinations \citep{moe_mind_2017}. To account for these uncertainties, we produce a set of model variations with a diverse set of assumptions on the primordial parameter distributions. 

Our fiducial model follows a Kroupa IMF distribution \citep{kroupa_variation_2001} for the stellar mass $m_{1}$ in the range [10,100] $\rm{M}_{\odot}$. This distribution is a power-law function with index $\alpha=-2.3$. Above a mass of $100\;\rm{M}_{\odot}$, extrapolation of the Hurley stellar evolution fitting formulae becomes less accurate. The stellar masses $m_{2}$ and $m_{3}$ are specified through the inner mass ratio $q_{\rm{in}} \equiv m_{2}/m_{1}$ and the outer mass ratio $q_{\rm{out}} \equiv m_{3}/(m_{1}+m_{2})$, respectively. Both the inner and outer mass ratios are sampled from a flat distribution in the range [0.1,1]\footnote{We have opted for a cut-off at $q=0.1$, as smaller mass ratios are rarely observed. However, the dearth of small mass ratios could be the result of observational selection effects. A low-mass  secondary or tertiary star is not expected to strongly impact the evolution of the primary star.}, based on mass-ratio observations of massive binary stars \citep{sana_binary_2012, kobulnicky_toward_2014}. The orbital semi-major axis range for $a_{\rm{in}}$ and $a_{\rm{out}}$ both cover [5,5$\times10^6$] $\rm{R}_{\odot}$, and the orbital distribution functions (see below) are sampled uniformly in logarithmic space, following \citep{opik1924statistical,kobulnicky_new_2007}. Below separations of $5\:\rm{R}_{\odot}$, the stars of the inner binary are likely to touch at the ZAMS. Above $5\times10^6\:\rm{R}_{\odot}$, the system is only very weakly bound through gravity and can easily be disrupted by small perturbations, such as stellar flybys \citep{kouwenhoven_primordial_2007, kouwenhoven_formation_2010}. When the sampling from our orbital semi-major axis distribution function yielded $a_{\rm{in}} > a_{\rm{out}}$, we decide to swap the values, as, by definition, $a_{\rm{in}} < a_{\rm{out}}$. This is a different approach from, e.g., one of the models in \citet{toonen_evolution_2020}, who fixed the primordial distribution of $a_{\rm{in}}$ and accepts or rejects $a_{\rm{out}}$ in accordance with the stability properties of the system. The resulting ZAMS distributions of $a_{\rm{in}}$ and $a_{\rm{out}}$ might differ between both methods. The eccentricities $e_{\rm{in}}$ and $e_{\rm{out}}$ are sampled from a thermal distribution in the range [0,0.9] \citep{ambartsumian_statistics_1937, heggie_binary_1975, moe_mind_2017, hwang_mystery_2023}. The arguments of pericenter $g_{\rm{in}}$ and $g_{\rm{out}}$ are sampled from a uniform distribution in the range [-$\pi$,$\pi$]. Lastly, the relative inclination $i_{\rm{rel}}$ is sampled from a uniform distribution in cosine in the range [0,$\pi$]. 

In each population model we vary only one (two in a single case) of the aforementioned distributions, in an effort to isolate the impact each parameter has on the evolution of the population. Five model variations assume the distributions for the semi-major axes and eccentricities as described by \citet{sana_binary_2012}, based on an observed population of O-type multiple stars in nearby galactic open clusters. The orbital periods are sampled from a power law in the range $0.15 < \rm{log}_{10}P < 8.5$, with index $\pi = -0.55$ and are then converted to semi-major axes using Kepler's third law. The eccentricities are sampled from a power law with index $\eta = -0.45$. In two model variations we assume a flat distribution of $e_{\rm{in}}$ and $e_{\rm{out}}$ \citep{kobulnicky_toward_2014}. In one variation, we assume a power law with index $\gamma=-1.5$ for the outer mass ratio, $q_{\rm{out}}$, which seems more appropriate for early B- and O-type primaries with wide companions \citep{moe_mind_2017}. For the final variation, we assume  $i_{\rm{rel}}=0$. This model is motivated by \citet{tokovinin_orbit_2017}, who showed that for outer orbital separations smaller than about $10^4\:\rm{R}_{\odot}$ the inner and outer orbits appear to be more aligned. Table \ref{table:init_param} provides a complete overview of the parameter distributions, sampling ranges and model variations.

\begin{table*}
\caption{The initial sampling ranges and distributions of the triple star properties for the fiducial model and all model variations.}             
\label{table:init_param}      
\centering                         
\begin{tabular}{l l l}        
\hline\hline   
& Fiducial model & \\
\hline\hline
sampling property & value/range & distribution  \\     
\hline\hline
  Initial primary mass $m_{1}$ & [10, 100] M$_{\odot}$ & Kroupa initial mass function \\
  Initial inner mass ratio $q_{\rm{in}}$ & [0.1, 1] & uniform distribution \\
  Initial outer mass ratio $q_{\rm{out}}$ & [0.1, 1] & uniform distribution \\
  Initial inner semi-major axis $a_{\rm{in}}$ & [5, 5$\times10^6$] R$_{\odot}$ & uniform distribution in log-space \\
  Initial outer semi-major axis $a_{\rm{out}}$ & [5, 5$\times10^6$] R$_{\odot}$ & uniform distribution in log-space \\
  Initial inner eccentricity $e_{\rm{in}}$ & [0, 0.9] & thermal distribution \\
  Initial outer eccentricity $e_{\rm{out}}$ & [0, 0.9] & thermal distribution \\
  Initial relative inclination $i_{\rm{rel}}$ & [0, $\pi$] & uniform distribution in cosine \\
  Initial inner argument of pericenter $g_{\rm{in}}$ & [$-\pi$, $\pi$] & uniform distribution \\
  Initial outer argument of pericenter $g_{\rm{out}}$ & [$-\pi$, $\pi$] & uniform distribution \\ 
  Initial metallicity $Z$ & Z$_{\odot}$ (0.014) & \\
  \hline\hline
& Model variations & \\ 
\hline\hline
name & varied parameter(s) & distribution \\
\hline\hline
  $a_{\rm{in}}$-Sana & $a_{\rm{in}}$ & powerlaw $(\rm{log}_{10}P_{\rm{in}})^\pi$, with $\pi = -0.55$. Based on \citet{sana_binary_2012} \\
  $a_{\rm{out}}$-Sana & $a_{\rm{out}}$ & powerlaw $(\rm{log}_{10}P_{\rm{out}})^\pi$, with $\pi = -0.55$. Based on \citet{sana_binary_2012} \\
  $a_{\rm{in}}\&a_{\rm{out}}$-Sana & $a_{\rm{in}}$ \& $a_{\rm{out}}$ & powerlaw $(\rm{log}_{10}P_{\rm{in\&out}})^\pi$, with $\pi = -0.55$. Based on \citet{sana_binary_2012} \\
  $e_{\rm{in}}$-Sana & $e_{\rm{in}}$ & powerlaw $e_{\rm{in}}^\eta$, with $\eta = -0.45$. Based on \citet{sana_binary_2012} \\
  $e_{\rm{out}}$-Sana & $e_{\rm{out}}$ & powerlaw $e_{\rm{in}}^\eta$, with $\eta = -0.45$. Based on \citet{sana_binary_2012} \\
  $e_{\rm{in}}$-flat & $e_{\rm{in}}$ & uniform distribution \\
  $e_{\rm{out}}$-flat & $e_{\rm{out}}$ & uniform distribution \\
  $q_{\rm{out}}$-Moe & $q_{\rm{out}}$ & powerlaw $q_{\rm{out}}^\gamma$, with $\gamma=-1.5$. Based on \citet{moe_mind_2017} \\
  $i_{\rm{rel}}$-const & $i_{\rm{rel}}$ & constant value of 0 \\

\hline\hline                                 
\end{tabular}
\end{table*}

\subsection{Dynamical stability}
\label{sect:stability}
To ensure long-term stability of a three-body system, a hierarchical configuration is generally required, i.e., the outer binary has a considerably larger orbit than the inner binary. In that case, the timescale at which the tertiary perturbs the system is long compared to the dynamical timescale of the inner binary. When the tertiary component orbits the inner binary too closely, the two timescales become comparable, breaking down the hierarchical structure and rendering the secular approximation invalid. We define the critical ratio for hierarchical stability following \citet{mardling_dynamics_1999, mardling_tidal_2001} as a criterion for hierarchical stability:

\begin{equation}
    \frac{a_{\rm{out}}}{a_{\rm{in}}}\Big\rvert_{\rm{crit}} = \frac{2.8}{1-e_{\rm{out}}}\Big(1-\frac{0.3i_{\rm{rel}}}{\pi}\Big)\Bigg[\frac{(1+q_{\rm{out}})(1+e_{\rm{out}})}{\sqrt{1-e_{\rm{out}}}}\Bigg]^{2/5}.
    \label{eq:instability}
\end{equation}
A system is dynamically stable if $a_{\rm{out}}/a_{\rm{in}} > (a_{\rm{out}}/a_{\rm{in}})_{\rm{crit}}$. For instance, for a circular orbit, equal mass system ($q_{\rm{out}}$=0.5), with the inner and outer orbit in the same plane, $(a_{\rm{out}}/a_{\rm{in}})_{\rm{crit}} = 3.3$. 

\begin{figure}[ht]
    \centering
    \includegraphics[width=\hsize]{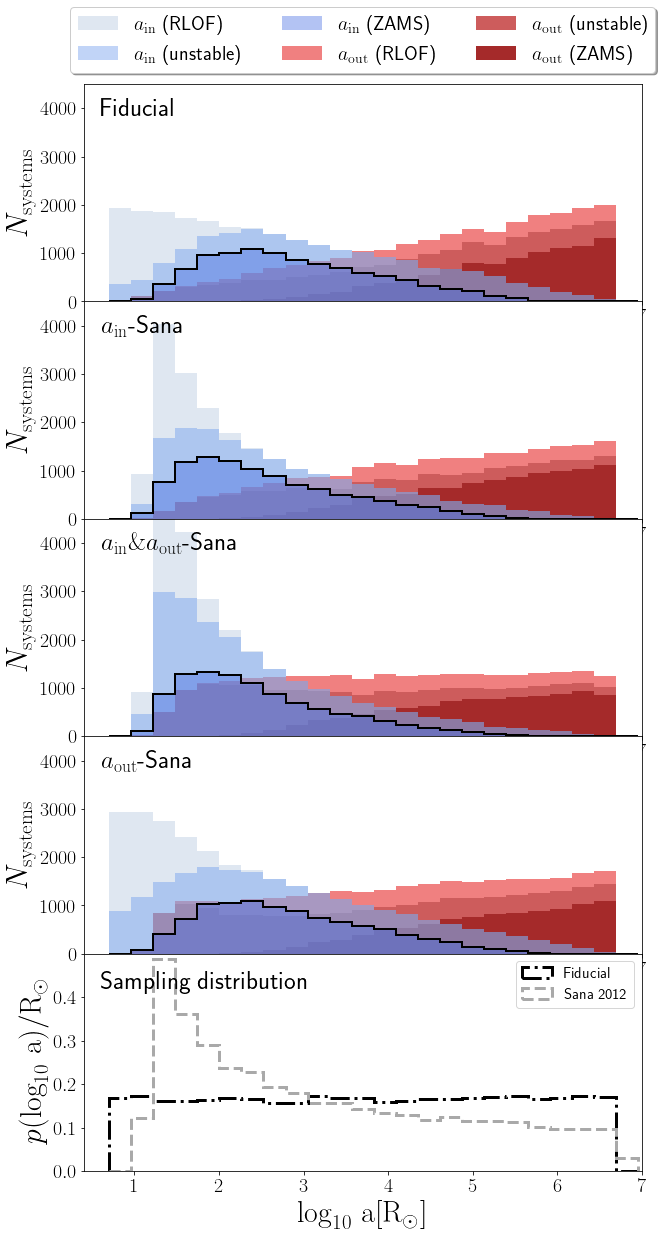}
    \caption{Initial distributions of the inner- (blue) and outer (red) semi-major axis for our fiducial model and three model variations. The solid black lines represent the ZAMS population, consisting exclusively of initially stable, non-interacting hierarchical triple stars. The lighter colours correspond to the population before excluding dynamically unstable systems and systems that are Roche-lobe filling at birth. To enhance the features in the plot, we have cut off the top part of the distribution for model $a_{\rm{in}\&\rm{out}}$-Sana. The bottom panel shows the initial sampling distributions of the fiducial model and the Sana distribution.}
    \label{fig:init_sep_comparison}
\end{figure}

Applying this stability criterion to our primordial population of triple stars affects the shape of the initial stellar and orbital parameter distributions. Fig. \ref{fig:init_sep_comparison} shows how the sampled distributions of $a_{\rm{in}}$ and $a_{\rm{out}}$ are altered by rejecting systems that are dynamically unstable or Roche-lobe filling at birth. While the fiducial model follows a flat sampling distribution in the logarithm of both $a_{\rm{in}}$ and $a_{\rm{out}}$, the population of hierarchical triple stars that satisfies the stability criterion (black line) is clearly more biased against large (small) inner (outer) semi-major axes. The most compact inner binaries ($a_{\rm{in}} < 20\:\rm{R}_{\odot}$) are often in contact at initialisation and are discarded. However, the number of systems discarded due to contact is much smaller for models with lower initial inner eccentricities. The important thing to notice is that the semi-major axis distributions of the ZAMS population of models that vary $a_{\rm{in}}$ and/or $a_{\rm{out}}$ are quite similar to the fiducial model, even though their sampling distributions are vastly differing.   

\subsection{Three-body dynamics}

The perturbations induced by the tertiary star onto the inner binary can be described through a mutual torque acting between the inner- and outer orbit. This torque allows angular momentum to be transported from one orbit to the other and vice versa, leading to an oscillatory behavior of the inner eccentricity and the relative inclination. The lowest order manifestation of this mechanism, the quadruple regime or ZLK effect, acts on a timescale that is mostly dependent on the ratio of orbital periods, $P^{2}_{\rm{out}}/P_{\rm{in}}$, and $e_{\rm{out}}$. An approximate expression for this timescale is \citep{kinoshita_analytical_1999, antognini_timescales_2015}:

\begin{equation}
    \tau_{\rm{ZLK}} \approx \frac{8}{15\pi}\Big(1+\frac{m_{1}}{m_{3}}\Big)\Big(\frac{P^{2}_{\rm{out}}}{P_{\rm{in}}}\Big)(1-e^{2}_{\rm{out}})^{3/2}.
    \label{eq:timescale}
\end{equation}
In the test-particle approximation, large oscillations of the inner eccentricity occur only at initial relative inclinations $i_{\rm{init}}$ between $39.2^\circ-140.8^\circ$ for initially non-eccentric inner orbits \citep{ford_secular_2000, naoz_secular_2013, naoz_eccentric_2016}. The maximum inner eccentricity reached through ZLK oscillations in the test-particle regime can be described analytically for circular orbits \citep{innanen_kozai_1997, grishin_generalized_2017}:

\begin{equation}
\label{eq:emax}
    e_{\rm{max}} = \sqrt{1-\frac{5}{3}\rm{cos}^{2}(\mathit{i}_{\rm{init}})}.
\end{equation}
Close to dynamical destabilisation and at non-zero outer eccentricities, higher order terms of the secular approximation treatment become important, such as the octupole term \citep{ford_secular_2000}. Besides reaching extreme eccentricity amplitudes, in this regime three-body dynamics can provoke a flip in relative inclination (from prograde to retrograde or vice versa) \citep{naoz_hot_2011, li_eccentricity_2014} and inner eccentricity variations can occur at a wider range of relative inclinations than $39.2^\circ-140.8^\circ$ \citep{ford_secular_2000}. The octupole term is relevant when the octupole parameter,

\begin{equation}
    \epsilon_{\rm{oct}} = \frac{m_{1}-m_{2}}{m_{1}+m_{2}}\frac{a_{\rm{in}}}{a_{\rm{out}}}\frac{e_{\rm{out}}}{1-e^{2}_{\rm{out}}},
\end{equation}
is of the order $|\epsilon_{\rm{oct}}| \gtrsim$ 0.001-0.01 \citep{lithwick_eccentric_2011, shappee_mass-loss-induced_2013}.

\subsubsection{Eccentricity oscillation tracking}

\begin{figure}[ht]
    \centering
    \includegraphics[width=\hsize]{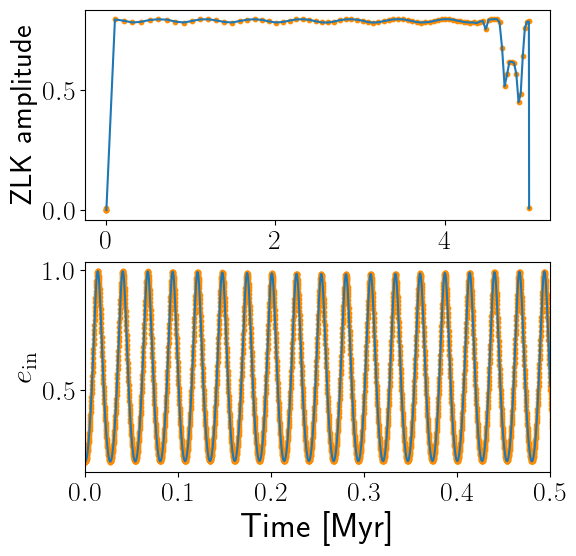}
    \caption{Example of a system with large inner eccentricity oscillations induced  by three-body dynamical effects. \textit{Top:} for each evolutionary time step, the maximum increase in the inner eccentricity is shown. During the first few time steps the ZLK amplitudes are small, because at the start of the simulation the evolutionary time steps are very short. \textit{Bottom:} the evolution of the eccentricity of the inner orbit during the first 0.5 Myr.}
    \label{fig:delta_e_in_time}
\end{figure}

We include a new function in the code that tracks the amplitude of the inner eccentricity oscillations during the entire evolution of a system. Per time step of the evolutionary code, the largest eccentricity amplitude is stored. This information can be used as a tool to quantify the impact of a third stellar component on the dynamical evolution of the system. In Fig. \ref{fig:delta_e_in_time} we show the maximum change in inner eccentricity during each time step of the evolutionary code for an example system. The system has initial stellar masses  $m_{1}=20\:\rm{M}_{\odot}$, $m_{2}=15\:\rm{M}_{\odot}$ and $m_{3}=30\:\rm{M}_{\odot}$, initial semi-major axes $a_{\rm{in}}=2\times10^4\:\rm{R}_{\odot}$ and $a_{\rm{out}}=1.5\times10^5\:\rm{R}_{\odot}$, initial eccentricities $e_{\rm{in}}=0.2$ and $e_{\rm{out}}=0.4$, and an initial relative inclination $i_{\rm{rel}}=\pi/2$. Until about 4.5 Myr, three-body dynamics persistently induce variations of the inner eccentricity up to 0.8. As the octupole parameter of this system is 0.009, the octupole term is important, which is manifested as the sinusoidal variations on timescales of about 0.8 Myr. With the analytical expression for the octupole timescale from \citet{antognini_timescales_2015}, we derive a comparable approximate oscillation time of 1.25 Myr. 
 
\subsection{Tidal interaction}

Stars in a binary system are subjected to tidal forces that act to synchronize the stellar rotation with the orbit and circularize the orbit by the dissipation of energy. Orbital properties, such as the semi-major axis, 
the eccentricity and the angular velocity are thereby affected and alter the further evolution and possibly final fate of the system. The tidal forces are strongly dependent on the ratio between the stellar radii and the semi-major axis, and hence only significantly impact compact binary systems. We adopt the equilibrium tidal model described by \citet{hut1980stability} for stars with convective envelopes. For stars with a radiative envelope, we adopt the dynamical tidal model described by \citet{zahn_tidal_1977}. Both the equilibrium and dynamical tidal models are parameterized by \citet{hurley_evolution_2002} and implemented in TRES. 

In orbital configurations where both the tidal dissipation and secular timescales are short compared to the evolutionary timescale of the stars, interplay between tides and three-body dynamics can efficiently reduce the semi-major axis of the inner binary \citep{mazeh_orbital_1979,kiseleva_tidal_1998}; ZLK oscillations increase the inner eccentricity, allowing tides to efficiently dissipate energy during pericenter passage \citep[e.g.][]{fabrycky_shrinking_2007}. After complete circularization, the semi-major axis has decreased by a factor $(1-e_{\rm{in}}^2)$.


\section{Results}


We briefly describe the conditions at which we stop individual stellar evolution simulations. We differentiate between the following channels:
\begin{itemize}
    \item [-] Mass transfer in the inner binary when either the primary or secondary star fills its Roche lobe.
    \item [-] Mass transfer from the tertiary star onto the inner binary, where the material may be accreted by one or both components.
    \item [-] Dynamical destabilisation due to stellar evolution (e.g., orbital widening due to mass loss) violating stability criterium Eq.~\ref{eq:instability}. 
    \item[-] Dynamical unbinding of the inner orbit when either the primary or secondary star experiences its core-collapse supernova.
    \item[-] Dynamical unbinding of the outer orbit as a result of core-collapse of one of the three system components.
    \item [-] None of the above interactions occur within a Hubble time. At this point all three components are compact remnants. We refer to this channel as non-interacting systems.
\end{itemize}
We do not explicitly identify the merging of two stars as a stopping condition as such events are preceded by mass transfer. Note that we do not include tidal effects and three-body dynamical effects such as ZLK oscillations as stopping conditions, while technically these could be described as orbital interactions. An overview of the evolutionary channels leading up to a model discontinuation is presented in Table \ref{table:ev_types}. 

The remainder of this section is structured as follows: in Section \ref{sect: overview channels} we give an overview of the predictions of the evolutionary channels for our simulated populations. Then we discuss two channels in more detail; the systems that experience a phase of mass transfer initiated by the primary star in Section \ref{sect: mass transfer primary} and the systems that do not undergo any interaction during their entire evolution in Section \ref{sect: non-interacting}. Next, we dive into few of the model variations in more detail in Section \ref{sect: model variations}. Finally, we compare the predictions for massive triple star evolution with simulations of isolated binary stars in Section \ref{sect: comparison binaries}.

\begin{table}
\caption{Overview and classification of the different evolutionary channels considered in this study.}           
\label{table:ev_types}      
\centering                         
\begin{tabular}{c c}        
\hline\hline    
\textbf{Interaction Type} & \textbf{Evolutionary Channel}  \\     
\hline\hline  
  Stellar Interaction & Inner mass transfer  \\
  & Tertiary mass transfer \\  \hline
  Orbital Interaction & Dynamical destabilisation  \\
  & Inner orbit unbound \\
  & Outer orbit unbound \\\hline
  Other & Non-Interacting \\
  
\hline\hline                                 
\end{tabular}
\end{table}

\subsection{Overview evolutionary channels} \label{sect: overview channels}

\begin{figure}[ht]
    \centering
    \includegraphics[width=\hsize]{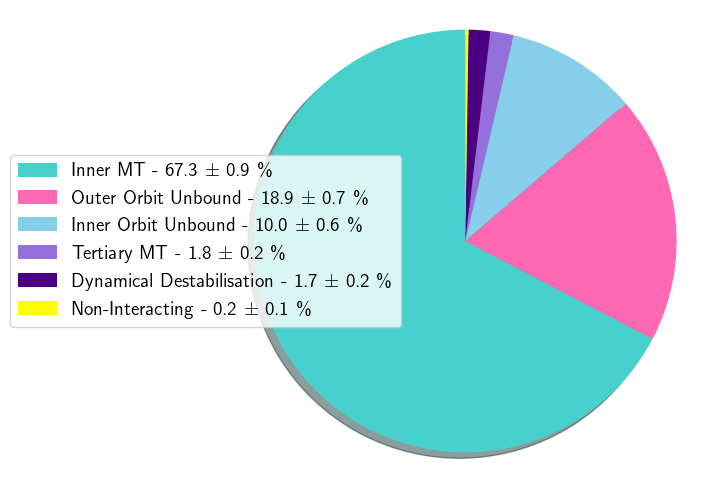}
    \caption{Percentage of systems evolving through each evolutionary channel for our fiducial population of massive hierarchical triple stars. For the outer orbit unbound, tertiary mass transfer and dynamical destabilisation channels, the tertiary star is directly involved.}
    \label{fig:pie_chart}
\end{figure}

In Fig. \ref{fig:pie_chart} we show the predicted contribution of each evolutionary channel to the total population of massive triple stars for our fiducial model. Similar to massive binary star systems \citep[e.g.][]{sana_binary_2012}, the evolution of massive triple star systems is dominated by both stellar and orbital interactions. The vast majority of systems ($67.3\pm0.9\%$) experiences an episode of mass transfer in the inner binary. This includes mass transfer initiated by both the primary and the secondary star, although the contribution of the latter is only very minor ($\lesssim$0.1\%). The error bars are the $3\sigma$ statistical sampling uncertainties obtained by bootstrapping the data (i.e., re-sampling the data with replacement). These uncertainties follow a Poisson distribution by approximation. A fairly large fraction of systems loses a companion as a result of a SN kick, unbinding either the inner- or outer orbit ($10.0\pm0.6\%$ and $18.9\pm0.7\%$ respectively)\footnote{We followed the SN kick model of \citet{verbunt_observed_2017} in combination with the fallback prescription of \citet{fryer_compact_2012} to scale down the kick velocities for black holes.}. In case the velocity kick is not strong enough to unbind the orbit, it can still affect the system by altering the semi-major axis and the eccentricity \citep{pijloo_asymmetric_2012, lu_supernovae_2019}. Only a few percent of the systems experiences mass transfer initiated by the tertiary star ($1.8\pm0.2\%$) or reach dynamical destabilisation ($1.7\pm0.2\%$). By far the smallest fraction of systems ($0.2\pm0.1\%$) has not engaged in one of the aforementioned interactions within a Hubble time. A detailed overview of the rates of all population models can be found in Appendix \ref{sect:appendix}.

\subsubsection{Model variations}

\begin{figure}
    \includegraphics[width=\hsize]{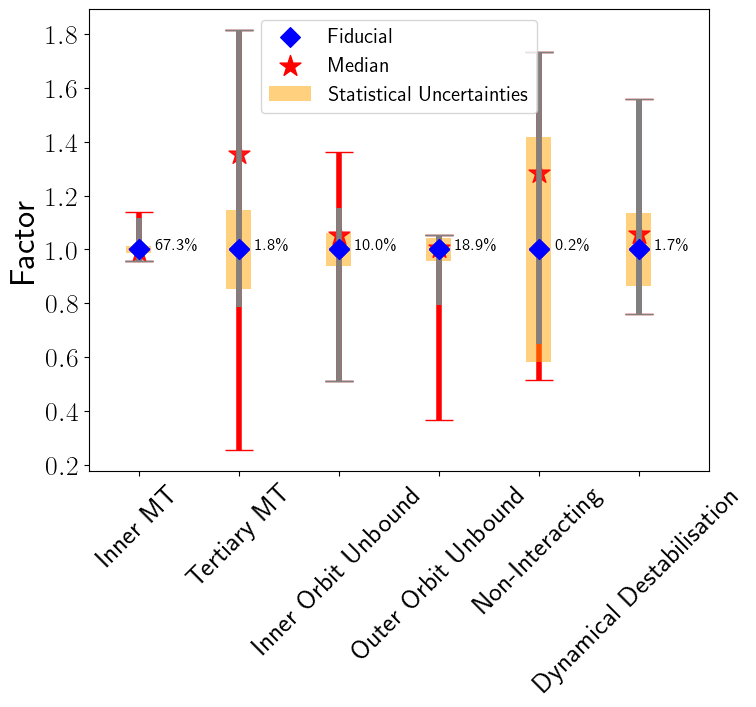}
    \caption{The predicted number of systems that evolve through each evolutionary channel for all model variations with respect to the fiducial model (blue diamonds). A factor 0.5 and 2 indicate a factor two increase and decrease, respectively. The predicted rates of all model variations are combined into the median (red stars), and the largest outliers per channel are represented by the error bars. The red part of the error bars correspond to Model $q_{\rm{out}}$-Moe, the model for which the predictions deviate most significantly from the fiducial model. The statistical uncertainties ($3\sigma$) on the predictions for the fiducial model are also included (orange bars). The relative contribution of each evolutionary channel for the fiducial are given in percentages.}
    \label{fig:population_comparison}
\end{figure}

Apart from statistical uncertainties, the predicted contributions of evolutionary channels are dependent on the uncertainties related to the formation of the stellar system, i.e. initial stellar and orbital parameters. In Fig. \ref{fig:population_comparison}, we show the predicted ranges on the contributing fraction of each evolutionary channel resulting from the model variations discussed in Section \ref{sect: init_params} relative to the fiducial model. Overall, the incidence rate of each evolutionary channel is fairly robust across the model variations; The rates only twice differ over a factor 2 compared to the fiducial model. The most significant outliers mainly originate from one model variation, Model $q_{\rm{out}}$-Moe. We discuss some of the outliers in more detail in Section \ref{sect: model variations}. Moreover, the predicted rates for the fiducial model are close to the median for most of the evolutionary channels, confirming that the fiducial model is a good representation of an average model. For the more uncommon channels, such as tertiary mass transfer, non-interacting and dynamically destabilised systems, the fiducial model predictions agree less well with the median. Although, we can not rule out that this deviation is explained by the enhanced statistical uncertainties for the latter two channels. 

In conclusion, the general evolution of a massive triple star population leading up to the first interaction is not seriously affected by the differing assumptions of the initial conditions. This can partially be explained by the initial stability check described in Section \ref{sect:stability}, reducing the differences in the ZAMS population between models with vastly differing initial parameter distributions. This is fortunate, as the initial parameters are empirically not well constrained. We stress that robust predictions for the incidence rates of evolutionary channels between two model variations do not necessarily imply that the complete evolution of stellar systems in both populations are identical. We terminate the simulation at the onset of the first interaction, but the outcome of certain interactions, such as mass transfer, and thus the subsequent evolution of the system, can be highly dependent on the stellar and orbital properties of the system at the ZAMS and similarly at the onset of the interaction. This is specifically important for the formation of more uncommon sources, such as gravitational wave mergers.

\begin{figure}
    \centering
    \includegraphics[width=\hsize]{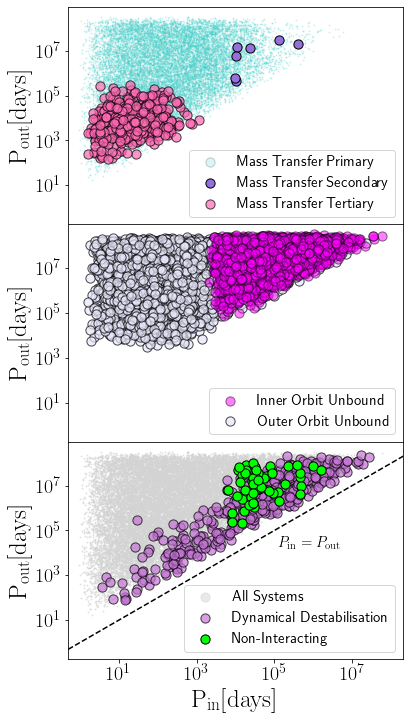}
    \caption{Initial inner- and outer orbital periods covered by each evolutionary channel. The triangle shape of the population is a direct consequence of the requirement on hierarchical structure and dynamical stability. The black dashed line shows where $P_{\rm{in}} = P_{\rm{out}}$. There are no systems near this line due to the dynamical stability criterion. There is some overlap between colors, as the evolution of a system is determined by more properties than just the orbital periods.}
    \label{fig:population_period}
\end{figure}

\subsubsection{Effect of initial conditions}
\label{sect:demography}
We provide an overview of a subset of the initial conditions that generally lead to a specific evolutionary channel in Fig. \ref{fig:population_period}, based on the triple systems in our fiducial model.
\begin{itemize}
    \item[--] \textit{Mass transfer:}
    Mass transfer initiated by the primary star can occur over a wide range of the inner and outer orbital period ($1-10^{6.5}\;\rm{days}$ for $\rm{P}_{\rm{in}}$ and $20-10^{8.5}\;\rm{days}$ for $\rm{P}_{\rm{out}}$), as the maximum radius of massive stars can be large. Even at $\rm{P}_{\rm{in}} > 10^4\;\rm{days}$, extreme inner eccentricities induced by ZLK oscillations can provoke the primary star to fill its Roche lobe during pericenter passage. For a few systems, the mass transfer in the inner binary is instead initiated by the secondary star. This can be counter-intuitive as the nuclear timescale of the primary star is shorter than that of the secondary star, hence the primary will evolve faster. A possible way to prevent the primary star from transferring mass first is by stripping the envelope via stellar winds before the star expands sufficiently to fill its Roche lobe. This does require the primary star to be very massive and is susceptible to the choice of wind model. Systems in which the secondary star transfers mass as their first interaction are restricted to initial inner orbital periods larger than $\gtrsim$10$^{4}$ days. As the triple systems are hierarchical, and thus the outer orbit is significantly wider than the inner orbit, the first phase of mass transfer will only be initiated by the tertiary star if it is initially more massive than either component of the inner binary. Therefore, the distribution of $q_{\rm{out}}$ is a key parameter in determining the contribution of this channel and $P_{\rm{in}}$ and $P_{\rm{out}}$ are restricted within about $10^{3}$ and $10^{5}-10^{6}$ days, respectively, for this channel. 
    \item[--] \textit{Orbital unbinding:} The initial conditions for systems whose inner (outer) orbit become unbound as a result of a SN kick are restricted to inner (outer) orbital periods larger than a few thousand days. Their subsequent evolution is difficult to predict without proper n-body calculations, as a phase of non-secular and possibly chaotic evolution ensues. However, it is not unlikely to expect at least one of the bodies to be ejected, either directly as a result of the SN kick, or through subsequent dynamical interactions \citep{pijloo_asymmetric_2012}.  
    \item[--] \textit{Dynamical destabilisation:} most systems that become dynamically unstable do not have very large initial ratios of $\rm{P}_{out}/\rm{P}_{in}$ (generally within a hundred, apart from a few exceptions). The location of instability is dependent on other parameters as well, such as $e_{\rm{out}}$, $q_{\rm{out}}$ and $i_{\rm{rel}}$, resulting in a quite generous spread of initial period ratios where systems experience a dynamical destabilisation. We identify several processes responsible for the transition from dynamically stable to unstable, the first being systems born close to the instability limit. Early on in their evolution, when the primary star is still on the main sequence (MS), three-body dynamical interactions can increase $e_{\rm{out}}$ or decrease $i_{\rm{rel}}$, shifting the critical semi-major axis ratio towards larger values \citep{toonen_evolution_2020}. Furthermore, once the primary star has evolved off the MS, either the increased mass-loss rate, resulting in the widening of the inner orbit, or an eventual SN kick, can push the system across the instability limit. 
    \item[--] \textit{Non-interacting:}
    Systems that do not have any form of interaction after a Hubble time of evolution are restricted to initial inner orbital periods of $\gtrsim6\times10^{3}$ days. Smaller periods would most likely result in a phase of mass transfer within the inner binary. 
\end{itemize}

\subsubsection{Time-evolution of evolutionary channels}

\begin{figure*}    \includegraphics[width=\textwidth]{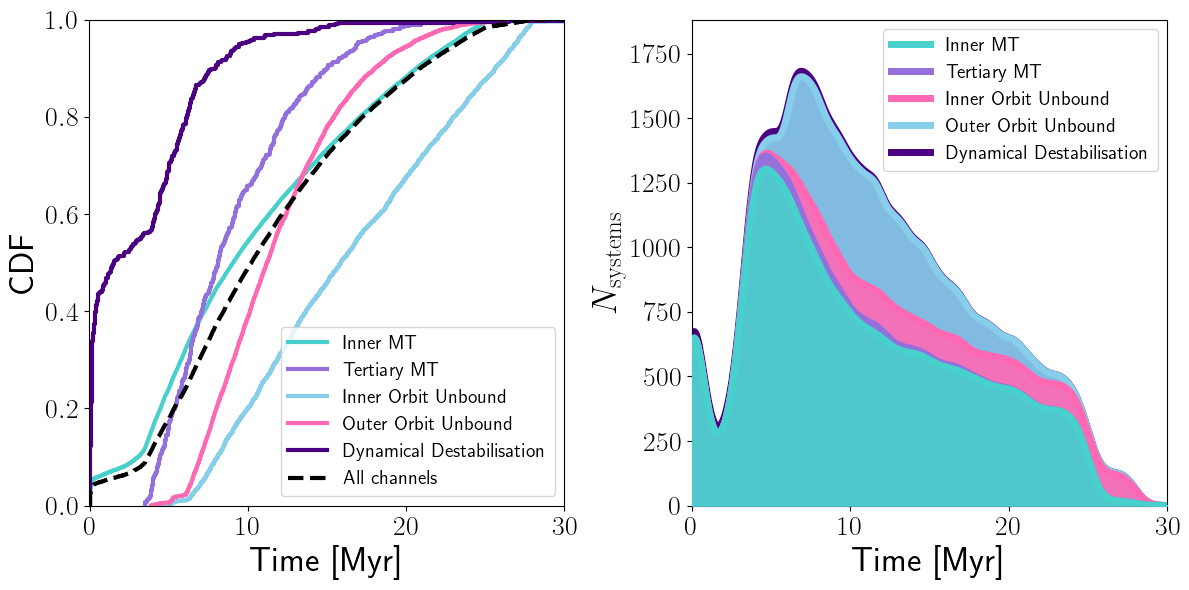}
    \caption{The rate of interactions as a function of time for the fiducial model. \textit{Left:} cumulative normalized distribution per evolutionary channel. The black dashed line combines the contribution of all channels. \textit{Right:} The total number of interactions occurring at each moment in time, summed across all channels.}
    \label{fig:channels_time}
\end{figure*}

The stellar and orbital properties of a system are affected when interactions between stars or orbits take place. Since interactions occur at different times for each system, the observational properties of a stellar population are expected to change over time. In Fig. \ref{fig:channels_time}, we present an overview of the typical stellar ages where each channel dominates. The non-interacting systems have been excluded, simply because their definition comprises the absence of any interaction. 

At very early times ($\lesssim4\:\rm{Myr}$), we predict solely dynamical destabilisation of the triple and mass transfer in the inner binary to take place. The extend of this epoch corresponds to roughly the MS lifetime of the most massive stars in our simulations. Both channels show a steep rise initially ($<1\:\rm{Myr}$), comprising systems that have short secular timescales and/or are initialised near the instability limit. Later on, the rate of interactions are small.

At later times ($\gtrsim4\:\rm{Myr}$), due to the fast radial expansion of post-MS stars and an eventual SN kick, the number of interactions increases rapidly within each channel and peaks just after 6.9 Myr. Interestingly, the systems in which the outer orbit becomes unbound typically interact at earlier times than systems in which the inner orbit becomes unbound. This may suggest that the former channel is biased towards high tertiary masses, as those stars evolve into a compact object on a shorter timescale. This is generally not the case. The explanation is that the inner binaries are generally more compact and have a higher binding energy compared to the systems in which the inner orbit becomes unbound. Therefore, for the inner binary the low kick velocities of the highest mass stars in our simulation are simply not sufficient to unbind their orbits. The rate of systems that undergo orbital unbinding of the outer binary or mass transfer from the tertiary onto the inner binary drops quickly after about 15 Myr. The reason for this is that $m_{1}$ has a sharp cut-off at $10\:\rm{M}_{\odot}$, while $m_{3}$ extends to masses as low as $1.2\:\rm{M}_{\odot}$. These low-mass stars do not experience a SN at the end of their lives, and reach different maximum radii compared to their massive counterparts. The exclusion of low-mass primary stars leads to a sharp drop in the number of systems that undergo a phase of mass transfer in the inner binary after 25 Myr. 

Overall, we predict that within 3.7 Myr about 10\% of systems have had an interaction. This fraction rapidly increase to $\sim$50\% after 10.2 Myr. After that, the rate of interactions slows down, despite the majority of stars evolving on timescales longer than 10 Myr as a consequence of the initial mass function. This trend follows the declining rate of inner binary mass-transfer systems, as that constitutes the most frequent type of interaction.

\begin{figure}
    \centering
    \includegraphics[width=\hsize]{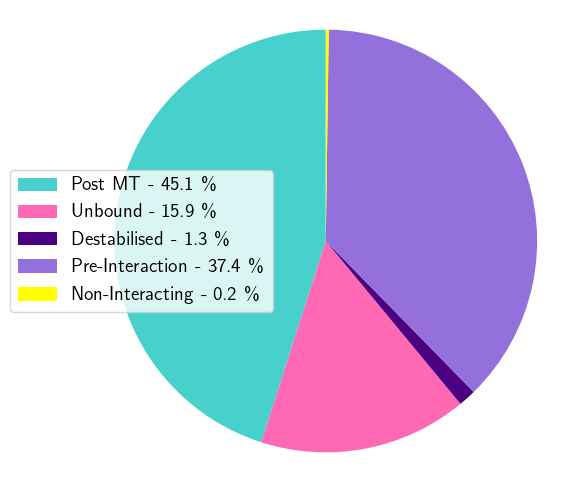}
    \caption{The incidence of pre- and post interacting massive triples assuming that stars formed at a constant star-formation rate. The contribution of binary stars are not included.}
    \label{fig:pie_chart_SFR}
\end{figure}

Lastly, we ask ourselves how the triple population looks like for a realistic stellar population. This can be done by convolving the results from Fig. \ref{fig:channels_time} with a star formation history. In Fig. \ref{fig:pie_chart_SFR}, we show how the intrinsic population of massive triples looks like after 30 Myr, which is the typical lifetime of star-forming giant molecular clouds, assuming a constant star-formation rate, rather than a single burst of star formation. The majority of systems (62.3\%) has already interacted. Most systems (45.1\%) have undergone a phase of mass transfer, initiated by any of the three stars. The high incidence of mass transfer is consistent with predictions for massive binary stars \citep{de_mink_incidence_2014} and emphasizes the importance of mass transfer in multiple stars. A significantly smaller fraction of the systems (15.9\%) has ejected at least one of the stars, while few systems (1.3\%) have destabilised as a result of three-body dynamics. \citet{de_mink_incidence_2014} showed that almost a third of post-mass-transfer binary systems result in a merger. Our findings could thus indicate that a large fraction of systems born as massive triple stars would be reduced to a population of single and binary stars at the moment of observation.

\subsection{Mass transfer initiated by primary} \label{sect: mass transfer primary}

\begin{figure}
    \centering
    \includegraphics[width=\hsize]{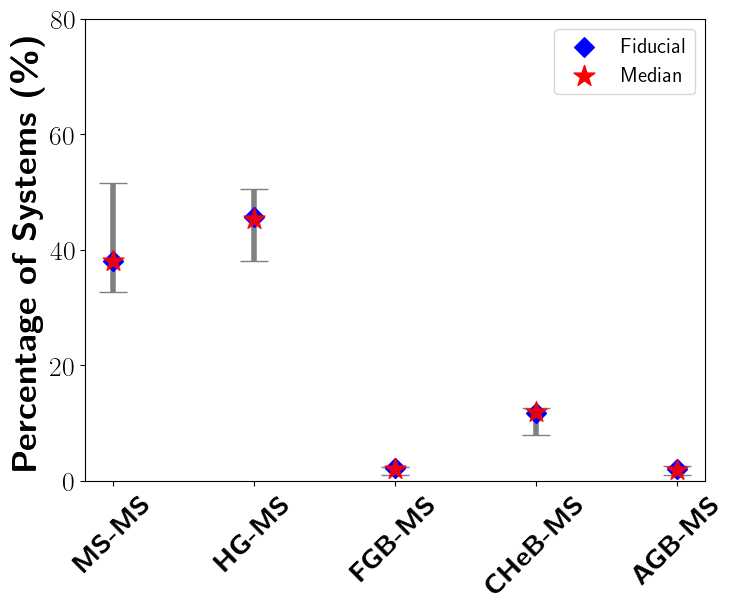}
    \caption{Evolutionary phase of the system at the onset of mass transfer in the inner binary. On the x-axis, the stellar types of the donor (primary) and accretor (secondary) are shown respectively. The symbols have the same meaning as those in Fig. \ref{fig:population_comparison}.}
    \label{fig:mt_comparison}
\end{figure}

The majority of systems (64.6-76.6\%) has a phase of mass transfer in the inner binary. As we stop our simulations at the first moment of interaction, the primary star will be the donor and the secondary star the accretor in most cases. One of the key ingredients in determining the outcome of mass-transfer systems is the stability of mass transfer. When a massive donor star approaches the giant branch, its envelope will transition from radiative to convective, and the mass transfer becomes more prone to instabilities \citep{klencki_it_2021}. Therefore, the evolutionary phase of the donor is a useful property to explore. We show the evolutionary phases of the donor and accretor stars at the onset of mass transfer for all model variations in Fig. \ref{fig:mt_comparison}. Given in chronological order, the phases at which the donor transfers mass are the Main Sequence (MS), Hertzsprung Gap (HG), First Giant Branch (FGB), Core Helium Burning (CHeB) and Asymptotic Giant Branch (AGB) phase. The CHeB phase is defined as the moment when helium is ignited in the core, which can occur before the star reaches the FGB phase for sufficiently high masses. These systems move directly from the HG phase to the CHeB phase, avoiding the FGB phase. We only discuss systems with a MS accretor, as the fraction of systems in which the secondary is an evolved star is extremely small ($\lesssim$0.5\%). Since the evolutionary phases after the MS are short lived compared to the total stellar lifetime, this can only occur when the donor and the accretor have similar evolutionary timescales, and thus similar initial masses. Nonetheless, these systems can lead to interesting events, such as a double core common envelope \citep{dewi_double-core_2006}.

In the vast majority of systems, the donor star is a MS star (32.8-51.7\%) or a HG star (37.8-50.5\%). Only 7.9-12.7\% of the donor stars are in the CHeB phase, while the FGB and AGB donors contribute merely a few percent. As discussed before, not all stars evolve through the FGB phase. Over 75\% of the primary stars are initially more massive than about 13M$_{\odot}$ and move directly to the CHeB phase. Similarly, stars initially more massive than about 38M$_{\odot}$ do not evolve through the AGB phase for our choice of wind mass-loss prescription and metallicity. Instead, they transition to the Wolf-Rayet phase while burning helium in the core. Additionally, for stars that do evolve through the AGB phase, the fractional increase in radius is generally marginal compared to the HG and CHeB phases.

We discuss two effects that explain the large number of systems that undergo mass transfer during an early evolutionary phase of the donor. First, as shown in Fig. \ref{fig:init_sep_comparison}, the ZAMS distribution peaks at small inner semi-major axes as a result of satisfying the criterion for dynamical stability, disfavouring large inner semi-major axes. At small separations, the donor star does not need to expand much to fill its Roche lobe, hence reaching the onset of mass transfer early in its evolution. Second, three-body dynamical interactions can expedite the onset of mass transfer by decreasing the pericenter of the inner binary through oscillations of the inner eccentricity. 

\begin{figure}
    \centering
    \includegraphics[width=\hsize]{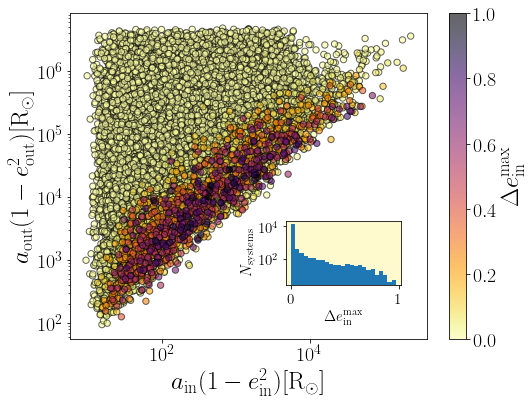}
    \caption{The circularization radius (the semi-major axis if the orbit was circularized by tides) of the inner and outer orbit of each system at the onset of mass transfer initiated by the primary star. Color-coded are the maximum amplitudes of the inner eccentricity induced by three-body dynamical interactions. Their distribution is also presented as a histogram. A value of $\Delta e^{\rm{max}}_{\rm{in}}=1$ indicates that during a single oscillation the inner eccentricity has varied between 0 and 1.}
    \label{fig:max_delta_ecc}
\end{figure}

To quantify the impact of three-body dynamics on the systems that go through a phase of mass transfer initiated by the primary star, we show the maximum amplitude of the eccentricity oscillations for each system of the fiducial model in Fig. \ref{fig:max_delta_ecc}. In 9.1\% of systems the inner eccentricity is increased between 0.05 and 0.96. The population model affected the most and the least by three-body dynamics are model $e_{\rm{in}}$-Sana, with low initial inner eccentricities, and model $i_{\rm{rel}}$-const, with initial relative inclinations of zero, respectively. For these models, 13.7\% and 1.1\% of the systems that go through a phase of mass transfer initiated by the primary star have experienced eccentricity oscillations with amplitudes between 0.05 and unity. The affected systems generally have small initial ratios of P$_{\rm{out}}$/P$_{\rm{in}}$, ensuring relatively short timescales on which the three-body dynamical interaction acts. However, there is a generous spread towards orbital period ratios up to over a hundred, since additional parameters ($e_{\rm{out}}$, $q_{\rm{out}}$ and $i_{\rm{rel}}$) affect the timescale and magnitude of the oscillations.

\subsection{Non-interacting systems} \label{sect: non-interacting}
Across all model variations only 0.1-0.4\% of the systems form a bound triple compact object (TCO) system that has had no previous interaction during their entire evolution. We have shown that these systems have initially relatively wide inner orbits to prevent a phase of mass transfer, and are therefore also required to have wide outer orbits to ensure a stable hierarchical configuration. At these wide separations however, the weakly bound tertiary star is prone to become unbound due to a SN kick, justifying the small contribution of this channel. 

Nevertheless, systems devoid of any interaction are an interesting target to study, as they are considered to be a main formation channel for producing gravitational wave mergers in hierarchical triple stars \citep{antonini_binary_2017, silsbee_lidovkozai_2017, rodriguez_triple_2018, fragione_black_2019, martinez_mass_2022}. In this scenario, the inner binary forms a double compact object (DCO) generally too wide to merge within a Hubble time. The tertiary star can drive the inner two objects significantly closer through three-body dynamics over long timescales, eventually merging the inner binary. Alternatively, in systems with a very wide outer orbit ($>1000\:\rm{AU}$), flybys can induce a gravitational wave merger by exciting the eccentricity of the outer orbit \citep{michaely_high_2020}. 

\begin{figure}
\centering
\includegraphics[width=\hsize]{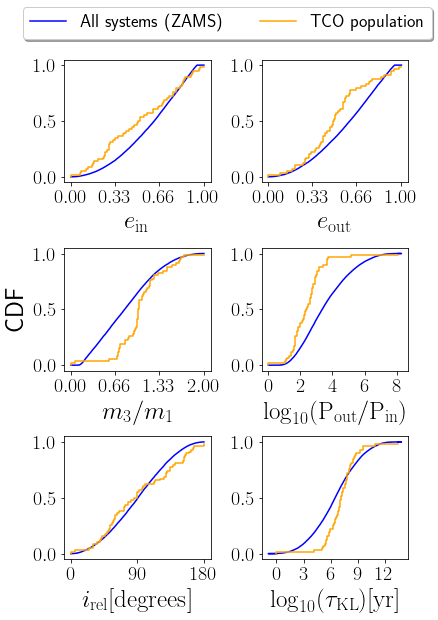}
  \caption{Cumulative step function of stellar and orbital properties for the fiducial model that are important in driving three-body dynamical interactions. In blue, we show the initial ZAMS conditions for all systems evolved in the simulation. In orange, the conditions at triple compact object formation are shown for the non-interacting systems.}
     \label{fig:non_interacting}
\end{figure}

In view of these possible compact object merger futures, we discuss the properties of the systems from the non-interacting channel. In Fig. \ref{fig:non_interacting}, we show the distribution of initial parameters that are of importance for driving the eccentricity oscillations. We compare these initial parameter distributions to the properties at TCO formation of the non-interacting systems for the fiducial model. Most of the properties are distributed quite dissimilar between the two populations. The TCO population evidently favors more moderate inner- and outer eccentricities. Highly eccentric systems are more likely to interact and hence prevent forming a TCO that has had no interaction during its evolution. Interestingly, the models $e_{\rm{in}}$-Sana and $e_{\rm{out}}$-Sana, that peak at much smaller initial inner- and outer eccentricities respectively, have eccentricities at TCO formation that are very similar to those in the fiducial model. For all model variations, near-equal mass ratios between the primary and tertiary compact object ($m_3/m_1$) are strongly favored. The system has to survive three subsequent SN kicks in order to form a bound TCO. For our choice of SN kick-velocity model, each star of the triple system needs to be massive to prevent break-up. Consequently, we predict a dearth of initial primary and tertiary masses below $35\:\rm{M}_{\odot}$. At high masses, all stars tend to evolve towards similar final black-hole masses due to high rates of wind mass loss. The orbital period ratios $\rm{P}_{\rm{out}}/\rm{P}_{\rm{in}}$ are almost completely limited to ratios smaller than $10^4$ for the TCO systems. The minimum inner orbital period to avoid mass transfer in the inner binary is around $6\times10^3$ days (see Fig. \ref{fig:population_period}). As the outer orbital periods are confined to $10^{8.5}$ days, the orbital period ratio never reaches extreme values. The TCO orbits slightly disfavor relative inclinations of $90^\circ$. At these inclinations, the eccentricity oscillations in the inner binary are the largest, complicating the formation of a non-interacting TCO.
The discussed properties set the timescale for the ZLK oscillations, which partially determines how effectively GW mergers can be produced due to the presence of a third compact object. The ZLK timescales (Eq. \ref{eq:timescale}) for the TCO systems are significantly longer than for the complete initial population and range from 100 kyr to 1 Gyr. However, ZLK oscillations and possibly higher order three-body dynamical interactions can still affect the system over the span of a Hubble time ($\sim13.5\:\rm{Gyr}$). 

\subsection{Differences in the incidence of evolutionary channels and onset of mass transfer between model variations} \label{sect: model variations}

The rates at which specific interactions occur vary quite significantly from the fiducial model for a few of the model variations (Fig. \ref{fig:population_comparison}). In particular, models that favor small inner semi-major axes,  such as model $a_{\rm{in}}$-Sana and $a_{\rm{in}}\&a_{\rm{out}}$-Sana, or low tertiary masses, such as model  $q_{\rm{out}}$-Moe, are responsible for the most significant differences. 

Model $a_{\rm{in}}$-Sana and $a_{\rm{in}}\&a_{\rm{out}}$-Sana show an increase in the incidence rate of systems that undergo mass transfer in the inner binary, with a rate of 73.4\% and 75.1\%, respectively, compared to the fiducial model's rate of $67.3\pm0.9\%$. These models have more compact inner orbits, resulting in smaller Roche radii of the inner binary components and thus increasing the likelihood of mass transfer. Model $a_{\rm{in}}\&a_{\rm{out}}$-Sana also has more systems where the tertiary star transfers mass onto the inner binary, with a rate of 3.2\%, compared to the fiducial model's rate of $1.8\pm0.2\%$. In this case, the outer orbits are more compact, making it more likely for the tertiary star to fill its Roche lobe.

Conversely, both models predict a decrease in the incidence rate of systems which unbound their inner orbit, with a rate of 5.9\% and 5.1\%, respectively. In the fiducial model, the inner orbits were unbound in $10.0\pm0.6\%$ of all systems. We identify two processes that contribute to this decrease. Firstly, the inner orbits are typically more compact, resulting in a larger binding energy, which means that the range of supernova kick velocities and directions required to unbind the orbit becomes smaller. Secondly, at smaller semi-major axes, the system is more likely to undergo mass transfer before the supernova occurs that ultimately unbinds the orbit.

Similar to models $a_{\rm{in}}$-Sana and $a_{\rm{in}}\&a_{\rm{out}}$-Sana, our predictions for model $q_{\rm{out}}$-Moe indicate an increase in the incidence rate of systems undergoing mass transfer in the inner binary, with a rate of 76.6\%. However, we also predict a significant decrease for systems transferring mass from the tertiary star onto the inner binary, with a rate of 0.45\%, compared to the fiducial model's rate of $1.8\pm0.2\%$. Additionally, we observe a large decrease in the incidence rate of systems experiencing an unbinding of the outer orbit. The predicted rate is 6.9\%, compared to the fiducial model's rate of $18.9\pm0.7\%$. These differences arise due to the fact that the population peaks at small initial outer mass ratios, resulting in lower tertiary masses. When the tertiary star is less massive than either component of the inner binary, its evolutionary timescale is longer, allowing the inner binary to initiate mass transfer or unbind the inner orbit via a supernova kick before the tertiary star has evolved sufficiently to interact. Moreover, a larger fraction of tertiary stars in model $q_{\rm{out}}$-Moe will not experience a SN kick at all, but will instead end their lives as white dwarfs.

For a range of models we predict a significant relative increase in the incidence rate of systems that do not interact, with values ranging from 0.38\% to 0.42\%, compared to the fiducial model's rate of 0.24\%. This includes all model variations with more moderate inner or outer eccentricities ($e_{\rm{in}}$-flat, $e_{\rm{out}}$-flat, $e_{\rm{in}}$-Sana \& $e_{\rm{out}}$-Sana) and a low relative inclination ($i_{\rm{rel}}$-const). At low eccentricities, the pericenter distance as well as the specific binding energy between stars is larger, making it less likely for the system to initiate mass transfer or unbind. At low relative inclination, three-body dynamical effects are less important and will not be able to induce eccentricity oscillations that drive the system toward interaction. 

In the models $a_{\rm{in}}$-Sana and $a_{\rm{in}}\&a_{\rm{out}}$-Sana, donors are usually less evolved, and the percentage of main sequence (MS) donors increases to 49.3\% and 51.7\%, respectively, compared to the fiducial model's 38.1\%. However, there are sligincludely fewer donors in later evolutionary phases, which is expected because a smaller inner semi-major axis results in a smaller Roche lobe for the donor star. For model $i_{\rm{rel}}$-const, we expect fewer MS donor stars and more HG donors. This is due to the fact that three-body dynamics are not effective at low relative inclinations, resulting in negligible inner eccentricity oscillations. Consequently, the pericenter of the inner binary does not decrease, and the onset of mass transfer is expedited.

\subsection{Comparison with isolated binaries} \label{sect: comparison binaries}

In this section we quantitatively investigate how the evolution of the inner binary is affected by a tertiary star. Previously, we have come across two mechanisms that could alter the evolution of a system with a tertiary component: (1) the initial ZAMS properties are affected by constraints of triple star formation and (2) three-body dynamical effects alter the orbital elements. In order to study the implications of both mechanisms on the evolution of the systems, we present two additional simulations of a population of isolated binary stars based on the same initial sampling distributions as the fiducial model. The first simulation represents a population of true isolated binary stars; i.e. the systems are initiated and evolved without the presence of a tertiary star. We call this model Bin. The systems in the second simulation are initiated as a hierarchical triple (identical to the fiducial model), forcing the initial distributions of the inner orbital parameters to become skewed to ensure a stable hierarchical triple configuration. Subsequently, the tertiary star is removed. Therefore, when we model the evolution of this system from the ZAMS onward, there are only two stars in the system. We call this model Bin-Skew. The inclusion of the latter model allows us to disentangle the contribution of tertiary induced dynamical effects from the selection effects introduced at star formation.

\begin{figure}
\centering
\includegraphics[width=\hsize]{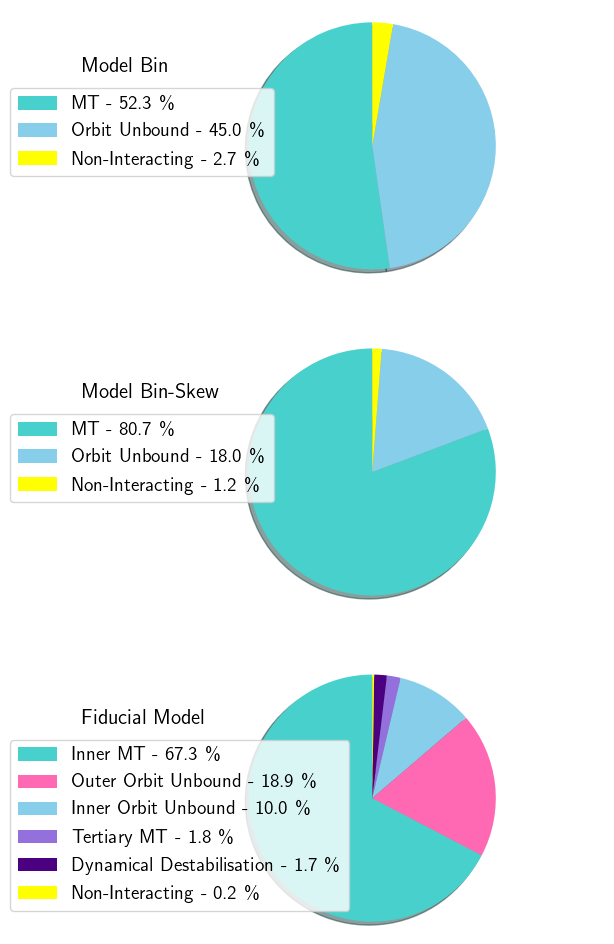}
  \caption{Percentage of systems evolving through each evolutionary channel for the isolated binary models Bin and Bin-Skew, and the fiducial model for reference. The systems in model Bin are initialised and evolved as isolated binary stars. The systems in model Bin-Skew are initialised as hierarchical triple stars, but evolved as isolated binary stars after removing the tertiary star. 
          }
     \label{fig:pie_chart_unskewed}
\end{figure}

We first investigate how significantly the general evolution of an isolated binary population is affected by skewing the orbital parameter distributions due to the presence of a tertiary star. To this end, we compare the predicted incidence rates of evolutionary channels for the models Bin and Bin-Skew (Fig. \ref{fig:pie_chart_unskewed}). There is an enormous discrepancy in the predicted number of systems that undergo mass transfer and systems whose orbit becomes unbound between both models. The model Bin-Skew has about one and a half times as many systems that experience an episode of mass transfer, 80.7\% compared to 52.3\%, and over a factor of two decrease in the number of orbits that become unbound, 18.0\% compared to 45.0\%. Furthermore, the number of systems that do not have any interaction during their entire evolution decreases by over a factor of two, 1.2\% compared to 2.7\%. The differences between these models are primarily a result of a smaller average pericenter at the ZAMS for the systems from model Bin-Skew. Interestingly, these differences are much bigger than the differences between the triple star models, as were shown in Fig. \ref{fig:population_comparison}. In short, dynamical constraints imposed by a third star during the initialisation of a system can alter the interaction history for a large number of systems. 

\begin{figure}
\centering
\includegraphics[width=\hsize]{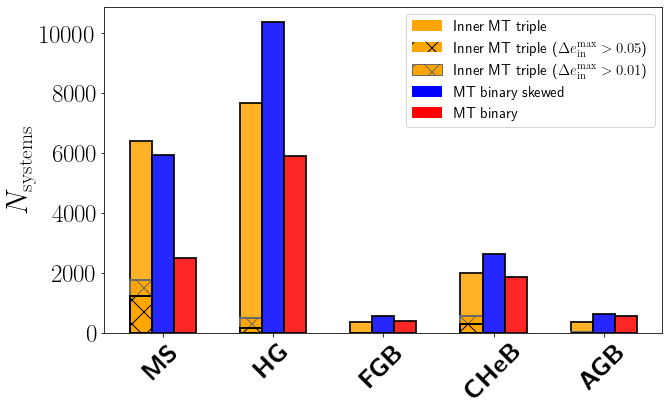}
  \caption{The number of systems per evolutionary phase that evolve through the (inner) mass transfer channel for the fiducial model (orange), the binary model with skewed initial parameters (blue) and the true isolated binary model (red). The hatched regions correspond to the number of inner binaries with maximum eccentricity amplitudes larger than 0.05 (black) and 0.01 (gray).
          }
     \label{fig:binary_vs_triple}
\end{figure}

Next, we investigate the effects of three-body dynamics on the evolution of the inner binary. To this end, we compare the occurrence of mass transfer from the primary onto the secondary star at different evolutionary phases of the donor star between the fiducial model and Bin-Skew (Fig. \ref{fig:binary_vs_triple}). For the fiducial model, we find an increase of 7.6\% in systems that initiate mass transfer on the MS compared to model Bin-Skew. Conversely, during later evolutionary stages, we predict a decrease of 23\%-46\%. Naively, one would have expected dynamical interactions to boost the total number of systems experiencing mass transfer in the inner binary. In only 67.3\% of the systems from the fiducial model the inner binary components undergo mass transfer, while for model Bin-Skew this is 80.7\%. This discrepancy follows from the fact that there are a few additional types of interactions unique to triple stars (Fig. \ref{fig:pie_chart_unskewed}). The systems in the fiducial model that undergo these unique triple-star interactions migincludegraphics still undergo mass transfer in the inner binary at a later stage of evolution, but their contribution is excluded here as the simulations are terminated at the onset of the first interaction. Therefore, a direct comparison between the predictions of the binary and triple simulations would be undesirable. 

However, the variations in the inner eccentricity due to three-body dynamics can give some insigincludegraphics into the difference in incidence rates of mass-transfer systems between the binary and triple simulations. 
For the triple population, the maximum oscillation amplitude of the inner eccentricity is most pronounced for the MS donors, with a measurable increase of more than 0.05 (0.01) in 19.1\% (27.6\%) of the systems. The MS lifetime is relatively long, which increases the opportunity for the quadruple terms as well as octupole terms to become relevant. At most other evolutionary phases the contribution of the tertiary star is significantly lower: 2.0\% (6.6\%) for HG donors, 0.27\% (1.9\%) for FGB donors, and 1.4\% (6.0\%) for AGB donors. The exception are CHeB donors, where the eccentricity increase is 15.4\% (28.0\%). Similar to the MS, the timescale of the CHeB phase is relatively long compared to the timescale of the other evolutionary phases. 

\section{Discussion}


\subsection{Comparison with other work}

\citet{hamers_double_2019} and \citet{stegmann_evolution_2022} both study the intermediate and final evolutionary stages of massive hierarchical triple stars in the field using the respective triple stellar evolution codes TSE and SECULARMULTIPLE \citep{hamers_secular_2016, hamers_secular_2018}. \citet{toonen_evolution_2020} addresses a similar scientific question as in this work, using the same code, but instead focus on the evolution of low mass triple stars ($1\:\rm{M}_{\odot} < m_{1} < 7.5\:\rm{M}_{\odot}$). We will compare the results of these studies with our results and comment on dissimilarities. 

\citet{hamers_double_2019} investigate the merger rates of double neutron star and black hole-neutron star mergers in hierarchical triple systems. They sample the initial stellar and orbital properties from similar distributions, but the sampling boundaries differ somewhat from our study. First of all, the initial primary masses are sampled between 8 and $50\:\rm{M}_{\odot}$, resulting in lower average masses. Second, their choice of the maximum initial inner orbital period is significantly longer compared to our study: $10^{10}$ days opposed to $10^{8.5}$ days. \citet{stegmann_evolution_2022} investigate the evolution of a massive triple star population from the ZAMS until the formation of a compact object. Similar to \citet{hamers_double_2019}, the initial primary masses reach down to $8\:\rm{M}_{\odot}$. However, instead of sampling each initial property from an independent distribution, they use correlated distributions as presented in \citet{moe_mind_2017}. Both studies do not vary the initial properties, but do vary physics, such as the supernova kick model. 

\subsubsection{Mass transfer initiated by primary}

\citet{hamers_double_2019} find that mass transfer initiated by the primary star occurs in 44\% of systems when including a non-zero SN kick prescription. That is 20\% fewer systems than the lower boundary on our predictions. The most likely explanation for this discrepancy is that their initial orbits extend to longer periods. Naturally, with less strongly bound orbits, we would expect a SN kick to unbind the system more easily. This is supported by an increase of over 20\% in the number of system where the inner and outer orbit becomes unbound in \citet{hamers_double_2019}. Furthermore, with lower typical masses, eventual SN kicks are stronger on average and are hence more likely to unbind the orbit. It is evident that changes in the  initial orbital parameter ranges can significantly alter the interaction history of a triple star population. 

\citet{stegmann_evolution_2022} find that at solar metallicity 74\% of the systems undergo mass transfer in the inner binary. However, since they do not terminate the simulation at the onset of an interaction, this population includes systems that experience the unbinding of the outer orbit prior to the mass transfer event. If the tertiary star was originally more massive than either component of the inner binary, mass transfer in the inner binary could be achieved even after the SN kick of the tertiary star has unbound the outer orbit. Unsurprisingly, our results indicate that for most models a lower fraction of inner binaries initiate mass transfer, with a median of 67\%. Since we do not have predictions for the number of systems that experience mass transfer in the inner binary after the first interaction, we refrain from drawing conclusions about the comparison between the incidence rates of mass-transfer systems. 

The frequency of mass transfer episodes that are initiated by the primary star in low-mass hierarchical triple populations is similar as in our high mass population \citep{toonen_evolution_2020}. However, the mass transfer is dominated by more evolved donor stars, namely by red giant branch stars, while in our simulations the mass transfer is mainly initiated by donor stars on the MS or HG. Opposed to massive stars, low mass stars barely expand during the MS phase and avoid filling their Roche lobe. Additionally, low-mass stars leave the MS phase much closer to the Hayashi track, resulting in a shorter Hertzsprung gap phase.

\subsubsection{Non-Interacting systems}
Previous studies have shown that the formation rate of TCO systems that have not experienced an interaction are highly dependent on the assumed SN kick model \citep{antonini_binary_2017, silsbee_lidovkozai_2017, rodriguez_triple_2018, fragione_black_2019}. With lower kick velocities, the triple orbits are more likely to remain bound and the probability of producing a TCO becomes higher. With non-zero kick velocities, \citet{stegmann_evolution_2022} and \citet{hamers_double_2019} predict a fraction of 0.12\% and 0.4\% non-interacting systems, respectively, which is within the range of our model predictions. However, when neglecting the SN kick, this fraction becomes substantially larger. Naively, one would expect that the compact-object merger rates would be drastically lower when higher kick velocities are implemented. Interestingly, \citet{antonini_binary_2017} showed that this is not the case for binary black-hole mergers, as bound systems that have received a stronger SN kick have larger eccentricities on average, making three-body dynamics more effective, which results in shorter merger delay times.

\subsubsection{GW sources}
Studies exploring the effect of three-body dynamics on the merger rate of compact objects in non-interacting triple stars have often neglected three-body dynamics prior to TCO formation. We explore the consequences of neglecting three-body dynamics during the stellar evolution phase on the population of non-interacting triple stars at TCO formation by excluding the quadrupole and octupole terms. Consistent with other studies, we still include the stability criterion. Our results indicate that neglecting three-body dynamics during the stellar evolution phase has an impact on the properties of the TCO population. For example, the fraction of systems with relative inclinations between $40^\circ$ and $140^\circ$ ($60^\circ$ and $120^\circ$) is 80\% (51\%) for the model variation compared to 70\% (34\%) for the fiducial model. The preference for aligned orbits is consistent across the other models. In addition, the fraction of systems in the model variation with outer eccentricities larger than 0.5 (0.6) is 59\% (46\%) compared to 45\% (27\%) for the fiducial population. However, not all other models show a clear preference towards lower outer eccentricities. The impact on the ZLK timescales is negligible, but the maximum amplitudes of the inner eccentricity due to three-body dynamical interactions are smaller for the fiducial population (Eq. \ref{eq:emax}). Specifically, 62\% of systems in the fiducial population have a maximum ZLK amplitude above 0.5, while this is 78\% for the model variation. This suggests that previous studies ignoring three-body dynamics during stellar evolution have overestimated the rate of gravitational wave mergers resulting from non-interacting triple stars. These results are not surprising, as stellar systems with strong dynamics are expected to interact before TCO formation. Furthermore, we predict no discernible difference in the final masses, mass ratios or orbital periods of the inner binary. 

\subsection{Main caveats}

Simulations with binary population synthesis codes rely on many assumptions that unavoidably introduce uncertainties to the predicted population. Adding a third star into the equation complicates matters further. In particular, the uncertainties in the initial orbital properties of the triples are prominent. Fortunately, these uncertainties are less important then one may expect (see Section 2.3). Besides the initial properties, we address the other main sources of uncertainty in the following section.

\subsubsection{Massive star evolution}
As mentioned in Section \ref{sect:method}, the approximate nature of the single stellar evolution fitting formulae can lead to imprecise inferred final properties of the stars, especially for stars above $50\rm{M}_{\odot}$, as this is the upper mass limit of the Hurley stellar tracks \citep{hurley_comprehensive_2000} and one needs to extrapolate at higher initial masses.

Furthermore, many aspects of stellar physics are still poorly understood. We highlight the two physical assumption most relevant for this study: mass loss throughout evolution and natal kick properties of compact objects. Uncertainties and intricacies in mass-loss from hot stars have recently been summarized by \cite{vink2022}. Scatter in empirical rates as function of luminosity are typically at least a factor of three, while discrepancies between theory and observations may exceed that amount in certain parameter ranges \citep[especially at luminosities below $\sim 10^{5}\,{\rm L}_{\odot}$; e.g.,][]{brands2022}. For cooler stars, uncertainties are even larger. For luminous stars that evolve toward the Humphrey-Davidson limit and experience a Luminous Blue Variable phase, accurate theoretical and observation-based mass-loss relations are lacking altogether \citep{smith_mass_2014}, forcing crude assumptions to be made. For Red Supergiants, which possibly suffer from multiple mechanisms causing the loss of gas \citep[e.g.,][]{cannon2021,montarges2021}, empirical prescriptions of mass loss deviate by up to an order of magnitude or more \citep[e.g.,][]{mauron2011,beasor2020}. These uncertainties are particularly concerning as stellar mass loss plays a crucial role in governing supernova progenitor properties, such as the core mass, and can affect the outcome of binary evolution. Higher wind mass-loss rates lead to more pronounced orbital widening and can prevent systems from engaging in mass transfer.

Of key relevance as well are the natal kick properties imparted on the compact object formed during a supernova event. We have assumed the kick velocity model from \citet{verbunt_observed_2017}, which is based on the motions of young galactic pulsars, and scaled down for black holes according to the fallback prescription of \citet{fryer_compact_2012} and the momentum of the object. However, the proper shape of the velocity distribution is still under debate. Some studies favour a single peaked Maxwellian distribution \citep{lyne_high_1994, hansen_pulsar_1997, hobbs_statistical_2005}, while others one that is double peaked \citep{fryer_population_1998, arzoumanian_velocity_2002, faucher-giguere_birth_2006, verbunt_observed_2017}. For black holes, the lack of observed natal kicks complicates discrimination between different models even further \citep{fragos_understanding_2009, repetto_investigating_2012, repetto_galactic_2017, mandel_estimates_2016}. The uncertainty in SN kicks on the occurrence rates of certain evolutionary channels can be significant (see Section 4.1.2). 

\subsubsection{Early interactions}
Roughly 4\% of the systems interact within $0.1\:\rm{Myr}$ from the ZAMS. These are mainly systems that engage in mass transfer and a small population of systems that become dynamically unstable. The vast majority of these systems has experienced an appreciable increase in the inner eccentricity, indicating three-body dynamical interactions have been at play. Only a handful of systems maintain a nearly constant inner eccentricity, as the systems were initialized extremely close to Roche-lobe filling or dynamical destabilisation. The number of systems that interact early on is on the order of 3-6\% across all model variations, apart from the model $i_{\rm{rel}}$-const. At initial relative inclinations of zero (coplanar orbits), we predict only 0.8\% of systems to interact within $0.1\:\rm{Myr}$, since at low inclinations three-body dynamics is not effective. 

We address a few points that could complicate the physical significance of these fast interactions. First, the initial inner semi-major axis of the systems experiencing mass transfer are typically not very large (few tens to few hundreds of solar radii). Before thermal equilibrium sets in on the ZAMS, the stellar radii are larger and the possibility can not be excluded that the components of the inner binary would interact on the pre-main sequence, likely resulting in a stellar merger \citep{tokovinin_formation_2020}. 

Second, observed low velocity dispersion distributions in young stellar clusters suggest that the minimum separation of inner binary orbits at formation might be larger than assumed in this study \citep{sana_dearth_2017, ramirez-tannus_relation_2021}, and that the Sana distribution is only established $\sim1\:\rm{Myr}$ after formation. With a larger minimum separation, we expect a decrease in the number of systems that interact early on, as systems with primary stars initially close to Roche-lobe filling are avoided and three-body dynamical effects need to be strong in order to drive wider systems to transfer mass. 

Third, observations of solar mass triple stars suggest that tight inner orbits, $a_{\rm{in}}<2000\:\rm{R}_{\odot}$, are associated with relative inclinations smaller than $40^{\circ}$ \citep{borkovits_comprehensive_2016,tokovinin_orbit_2017}. At low inclinations, three-body dynamical interactions are strongly reduced and likely diminish the contribution of these fast interactions. 
 
\subsubsection{Eccentric mass transfer}

\begin{figure}
\centering
\includegraphics[width=\hsize]{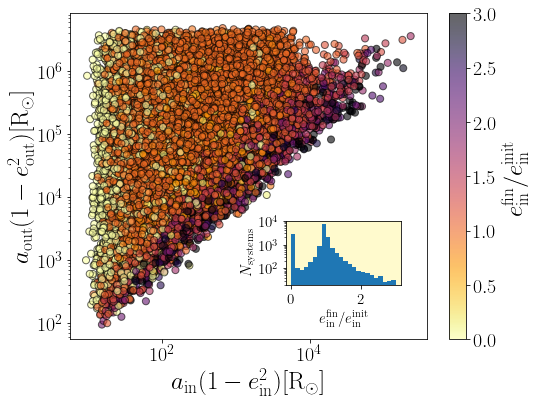}
  \caption{The circularization radius of the inner and outer orbit of each system at the onset of mass transfer initiated by the primary star. Color-coded is the ratio between the final and initial eccentricity of the inner orbit. Their distribution is also presented as a histogram in the smaller plot. The light-colored systems ($e^{\rm{fin}}_{\rm{in}}/e^{\rm{init}}_{\rm{in}} \ll 1$) at small circularization radii of the inner orbit have been completely circularized due to tidal dissipation. The dark-colored systems ($e^{\rm{fin}}_{\rm{in}}/e^{\rm{init}}_{\rm{in}} > 1$) at comparable inner- and outer circularization radii have been subjected to three-body dynamical effects. The evolution of the other systems has hardly been affected by either tidal effects or three-body dynamics.  
          }
     \label{fig:fin_vs_init_ecc}
\end{figure}

An important challenge in population synthesis codes is the lack of understanding of how mass transfer progresses in eccentric orbits. Usually, this problem is circumvented by assuming the system circularises by the onset of mass transfer or initialising the orbits as circular. However, recent studies find that the tidal forces are generally inefficient in circularising the orbit of binary systems before mass transfer is initiated \citep{eldridge_new-age_2009,vigna-gomez_common_2020, vick_tidal_2021}. Moreover, in triple star systems, if the three-body dynamical interactions are strong enough, complete circularisation due to tidal forces can be prevented \citep[e.g.][]{toonen_evolution_2020}. In Fig. \ref{fig:fin_vs_init_ecc}, we show the degree of circularisation of the inner orbit for all systems in which the primary star fills its Roche lobe for our fiducial model. While many of the most compact inner orbits are completely circularised, the majority of systems retain a nearly constant eccentricity. A non-negligible fraction of systems, located at small $a_{\rm{out}}/a_{\rm{in}}$ even experiences a significant boost in eccentricity as three-body dynamical effects become more important. We stress that Fig. \ref{fig:fin_vs_init_ecc} is based on a singular tidal model, where (1) the tidal timescales are uncertain and (2) tidal processes, such as efficient dissipation in highly eccentric orbits \citep{moe_dynamical_2018,generozov_overabundance_2018}, are ignored. As a result, we are likely to underestimate the degree and timescale of circularisation in the inner binary. This is especially valid for inner orbits with high eccentricities, but also for small semi-major axes, where tides could be more efficient during the pre-main sequence phase.

\subsubsection{Tertiary tides}

Besides tidal forces between the primary and secondary star, the tertiary star can also dissipate energy from the inner binary through tidal interactions \citep{fuller_tidally_2013}. To investigate the importance of this effect, we apply the tertiary tidal model of \citet{gao_empirical_2020} on our fiducial population. As a result of tertiary tides, the semi-major axis of the inner binary shrinks as:

 \begin{equation}
 \begin{aligned}
    \frac{1}{a_{\rm{in}}}\frac{da_{\rm{in}}}{dt} = 2.22\times10^{-8}\rm{yr}^{-1} \: \frac{4q_{\rm{in}}}{(1+q_{\rm{in}})^2} \: \Bigg(\frac{R_3}{100\:\rm{R}_{\odot}}\Bigg)^{5.2}\\
    \; \times \; \Big(\frac{a_{\rm{in}}}{0.2\:\rm{AU}}\Big)^{4.8} \: \Big(\frac{a_{\rm{out}}}{2\:\rm{AU}}\Big)^{-10.2} \: \Bigg(\frac{\tau}{0.534\:\rm{yr}}\Bigg)^{-1.0},
\end{aligned}
\end{equation}
with $R_{3}$ the radius and $\tau$ the viscoelastic relaxation time of the tertiary star. We set $\tau$ to $10^{-4}\:\rm{yr}$ in accordance with \citet{gao_empirical_2020}. We did not find a noticeable impact of the tertiary tides on the evolution of our fiducial population for the chosen value of $\tau$.


\section{Conclusions}

In this work, we have studied the main evolutionary pathways of massive hierarchical triple stars up to the first moment of interaction. Additionally, we have investigated what impact a tertiary companion has on the evolution of the system. We have done this by performing large-scale simulations of massive triple star evolution, beginning at the ZAMS up to the first point of stellar or orbital interaction. To account for uncertain formation properties in the simulated population, we included predictions for several model variations of the initial parameter distributions. Our key findings can be summarized as follows:

\begin{itemize}
    \item The initial parameters of stable hierarchical triple stars are highly constrained by the combined requirement of dynamical stability and avoidance of contact of the inner binary, which restrict the period distribution. 
    Triple initiation results in less wide inner orbits, leading to a high incidence of interactions such as mass transfer. The inner orbit period distribution strongly differs from a flat distribution and deviates substantially different from a Sana distribution.  
    
    \item The vast majority of systems (65\%-77\%) have a phase of mass transfer initiated by the primary star as their first interaction. This occurs mainly at early evolutionary stages of the donor. In 32-50\% of the cases the donor is still a main-sequence star, and it has evolved to the Hertzsprung gap in 38-50\% of the systems. We have identified two main processes responsible for the large number of systems that initiate mass transfer at an early evolutionary phase. First, as a result of the dynamical stability criterion at formation, the inner orbits are typically more compact initially, and therefore the stars need to expand less in order to fill their Roche lobe, thus favoring mass transfer. Second, three-body dynamics can drive up the maximum eccentricity of the inner orbit, shrinking the eccentric Roche lobe. In the fiducial model, 10\% of all systems that undergo mass transfer initiated by the primary star experience an increase of the inner eccentricity by at least 0.05. 
    
    \item Across all triple models, we predict that fewer than 0.5\% of the systems do not engage in any interaction after being evolved for a Hubble time, such that all three stars effectively evolve as single stars. These systems are known targets as progenitors for gravitational wave mergers. We have shown that systems with strong three-body dynamics tend to interact before a triple compact object (TCO) is formed and hence reduce the likelihood of producing compact objects that merge with the help of von Zeipel-Lidov-Kozai (ZLK) oscillations. However, the typical ZLK timescales are still a few orders of magnitude shorter than a Hubble time and can accelerate the inspiral of the compact objects. We have also shown that ignoring three-body dynamics before compact object formation results in TCOs with stronger eccentricity oscillations and thereby likely over-predicts the merger rate of compact objects in such systems.
    
    \item The predicted incidence rates of systems evolving through each evolutionary channel is most sensitive to variations in the inner/outer semi-major axis and the outer mass ratio distribution. The incidence rate of inner mass-transfer systems is dominant compared to other evolutionary channels leading to interaction. Though, the absolute differences between population models for these less common channels are not important, the relative differences can be relevant. For only two evolutionary channels, mass transfer from the tertiary star onto the inner binary and the unbinding of the outer orbit, the incidence rate differs by over a factor of two from the fiducial model's rate (see Fig. \ref{fig:population_comparison}).

    \item The evolution of an isolated massive binary star population up to the first interaction differs significantly from that of massive hierarchical triple stars. The criterion of dynamical stability and eccentricity oscillations due to dynamical interaction ensure that at least an additional 15\% of the population initiates mass transfer in the inner binary. Moreover, the donor stars are generally at an earlier evolutionary phase at the onset of mass transfer. 
 
\end{itemize}

\begin{acknowledgements}
The authors acknowledge support from the Netherlands Research Council NWO (VENI 639.041.645 and VIDI 203.061 grants).
\end{acknowledgements}

%
\bibliographystyle{aa} 
\bibliography{references} 
%


   

\begin{appendix} 

\section{Overview incidence rates} 
\label{sect:appendix}

\begin{table}[h!]
\caption{Normalised incidence rates of evolutionary channels for all population models. The $3\sigma$ errors are based on Poisson uncertainties obtained by bootstrapping of the data}           
\label{table:incidence_rates}      
\centering                         
\resizebox{\textwidth}{!}{\begin{tabular}{l c c c c c c}        
\hline\hline  
& Inner MT & Tertiary MT & Inner Orbit Unbound & Outer Orbit Unbound & Non-Interacting & Dynamical Destabilisation \\
\hline\hline
  Fiducial & $0.673\pm0.009$ & $0.018\pm0.003$ & $0.100\pm0.006$ & $0.189\pm0.008$ & $0.002\pm0.001$ & $0.017\pm0.002$ \\ 
  $a_{\rm{in}}$-Sana & $0.734\pm0.009$ & $0.024\pm0.003$ & $0.059\pm0.004$ & $0.168\pm0.007$ & $0.002\pm0.001$ & $0.013\pm0.002$ \\
  $a_{\rm{out}}$-Sana & $0.687\pm0.009$ & $0.024\pm0.003$ & $0.088\pm0.005$ & $0.180\pm0.007$ & $0.002\pm0.001$ & $0.019\pm0.003$ \\
  $a_{\rm{in}}\&a_{\rm{out}}$-Sana & $0.751\pm0.008$ & $0.032\pm0.003$ & $0.051\pm0.004$ & $0.150\pm0.007$ & $0.004\pm0.001$ & $0.016\pm0.002$\\
  $e_{\rm{in}}$-Sana & $0.644\pm0.009$ & $0.026\pm0.003$ & $0.115\pm0.006$ & $0.196\pm0.007$ & $0.004\pm0.001$ & $0.016\pm0.002$\\
  $e_{\rm{out}}$-Sana & $0.646\pm0.009$ & $0.014\pm0.002$ & $0.116\pm0.006$ & $0.195\pm0.008$ & $0.004\pm0.001$ & $0.026\pm0.003$\\
  $e_{\rm{in}}$-flat & $0.661\pm0.009$ & $0.025\pm0.003$ & $0.105\pm0.006$ & $0.192\pm0.007$ & $0.004\pm0.001$ & $0.015\pm0.002$\\
  $e_{\rm{out}}$-flat & $0.659\pm0.009$ & $0.015\pm0.002$ & $0.107\pm0.006$ & $0.192\pm0.008$ & $0.004\pm0.001$ & $0.023\pm0.003$\\
  $q_{\rm{out}}$-Moe & $0.766\pm0.008$ & $0.004\pm0.001$ & $0.137\pm0.007$ & $0.069\pm0.005$ & $0.001\pm0.001$ & $0.022\pm0.003$\\
  $i_{\rm{rel}}$-const & $0.646\pm0.009$ & $0.025\pm0.003$ & $0.107\pm0.006$ & $0.200\pm0.007$ & $0.004\pm0.001$ & $0.018\pm0.003$\\
  
\hline\hline                                 
\end{tabular}}
\end{table}

\end{appendix}

\end{document}